\documentclass[prl,twocolumn,showpacs,preprintnumbers,amsmath,amssymb]{revtex4}

\usepackage{graphicx}
\usepackage{dcolumn}
\usepackage{bm}
\usepackage{wrapfig} 
\usepackage{multirow}
\usepackage{color}

\newcommand{\beq}{\begin{eqnarray}} 
\newcommand{\eeq}{\end{eqnarray}}

\begin{document}

\preprint{CERN--PH--TH/2013--283}\vspace*{0.5cm}

\title{Complementarity of WIMP Sensitivity\\ with direct SUSY, Monojet and
Dark Matter Searches in the MSSM}

\author{Alexandre Arbey}
\email{alexandre.arbey@ens-lyon.fr}
\affiliation{Centre de Recherche Astrophysique de Lyon, Observatoire de Lyon,\\
\mbox{Saint-Genis Laval Cedex, F-69561, France; CNRS, UMR 5574;}\\ 
Ecole Normale Sup\'erieure de Lyon, France;\\
\mbox{Universit\'e de Lyon, Universit\'e Lyon 1, F-69622~Villeurbanne Cedex, France}\\
\mbox{and CERN, CH-1211 Geneva, Switzerland}}%

\author{Marco Battaglia}
\email{MBattaglia@lbl.gov}
\affiliation{University of California at Santa Cruz, \\
Santa Cruz Institute of Particle Physics, CA 95064, USA\\
and CERN, CH-1211 Geneva, Switzerland}%

\author{Farvah Mahmoudi}%
\email{mahmoudi@in2p3.fr}
\affiliation{Clermont Universit\'e, Universit\'e Blaise Pascal, CNRS/IN2P3,\\
LPC, BP 10448, F-63000 Clermont-Ferrand, France\\
and CERN, CH-1211 Geneva, Switzerland}%


\begin{abstract}
This letter presents new results on the combined sensitivity of the LHC and underground dark matter search 
experiments to the lightest neutralino as WIMP candidate in the minimal Supersymmetric extension of the Standard Model. 
We show that monojet searches significantly extend the sensitivity to the neutralino mass in scenarios where 
scalar quarks are nearly degenerate in mass with it. The inclusion of the latest bound by the LUX experiment 
on the neutralino-nucleon spin-independent scattering cross section expands this sensitivity further, highlighting the 
remarkable complementarity of jets/$\ell$s+MET and monojet at LHC and dark matter searches in probing models of new 
physics with a dark matter candidate. The qualitative results of our study remain valid after accounting for theoretical 
uncertainties.
\end{abstract}

\maketitle


Astrophysical data has convincingly established the existence of non-baryonic dark matter (DM), most likely due to a new kind of neutral, stable, 
weakly-interacting  massive particle (WIMP)~\cite{Bertone:2004pz}. 
One of the most appealing features of Supersymmetry with conserved R-parity (SUSY) is that it provides us with a natural candidate for 
WIMP dark matter~\cite{Pagels:1981ke,Weinberg:1982zq,Weinberg:1982tp,Ellis:1983ew,Jungman:1995df}. If the lightest supersymmetric particle (LSP) in 
the theory is the lightest neutralino, $\tilde{\chi}^0_1$, it represents a WIMP with the appropriate properties to account for the observed cosmic 
dark matter. A $\tilde{\chi}^0_1$ with mass of $\cal{O}$(100~GeV) and typical weak-interaction couplings, develops a relic density in the Universe 
of the same order as the value now precisely measured by satellite experiments from the analysis of the cosmic microwave background (CMB) 
spectrum~\cite{Hinshaw:2012aka,Ade:2013zuv}. The search for WIMPs has become one of the most compelling research areas at the intersection between 
collider physics, non-accelerator particle physics and cosmology~\cite{Baltz:2006fm}.
In fact, a WIMP neutralino can be searched for at the LHC, first in direct SUSY searches, from the decay chains of heavier SUSY particles through 
topologies with the production of Standard Model (SM) particles and significant missing transverse energy (MET), then in direct 
$\tilde{\chi} \tilde{\chi}$ pair production through ``monojet'' events. 
Direct DM searches can reveal signals of cosmic WIMP neutralinos through the small energy released in their scattering 
on the nucleons of the sensitive detector volume in underground experiments. 
All these searches are characterised by different sensitivities, depending on the SUSY parameters.
Within a well-defined model, these sensitivities can be compared and their complementarity and redundancy studied. 
In the study of WIMP neutralinos, the phenomenological minimal supersymmetric extension of the SM (pMSSM), a minimal, R-parity conserving SUSY model 
with 19 free parameters~\cite{Djouadi:1998di}, represents a very practical model providing a broad and unbiased perspective on the generic MSSM 
phenomenology. Its parametrisation ensures that the SUSY particle masses are independent, while keeping a small enough number of parameters limited to 
allow full scans of the model phase space to be performed.  
A comparison of the jets+MET and monojet searches at the LHC was already presented in~\cite{Dreiner:2012sh}, where the 
7~TeV data and specific masses were considered. An earlier general study of neutralinos as DM candidate in the pMSSM was discussed 
in~\cite{Cahill-Rowley:2013dpa}, specific pMSSM scenarios with a light WIMP neutralino in~\cite{Arbey:2012na,Arbey:2013aba}. 
In this letter, we report the results of the first study of WIMP neutralino sensitivity in the framework of the pMSSM, which includes the LHC 
jets/leptons+MET and the important monojet analyses on full 8~TeV data, the PLANCK CMB constraints and the new results on direct DM
searches from the LUX experiment. We show how the inclusion of the LHC monojet analyses and the DM direct 
detection bounds significantly expands the region of MSSM parameters probed by the data.  These searches can test regions of the parameter space 
not accessible to the LHC direct SUSY searches, characterised by specific kinematics of signal events, in particular with SUSY particles having 
masses almost degenerate with the LSP $\tilde{\chi}$. We develop a new methodology to interpret the monojet search results in 
SUSY in a largely model-independent way. The complementarity with jet+MET direct SUSY searches at the LHC and direct dark matter experiments proves 
to be remarkable. 
These conclusions hold also when we account for QCD uncertainties in the monojet event cross section and astrophysical uncertainties in the 
derivation of the $\tilde{\chi}$-$p$ scattering cross section, $\sigma_{\tilde{\chi} p}^{SI}$.    


This study is based on a large statistics scan of the pMSSM parameters, constraints from lower energy data and the determination of the sensitivity 
of various search analysis to the spectrum and decay pattern of each accepted pMSSM point. The tools used to perform the scans and the analysis 
have been presented in~\cite{Arbey:2011un,Arbey:2011aa}.
SUSY particle spectra are calculated using SOFTSUSY 3.2.3~\cite{softsusy}, the decay branching fractions for the Higgs and SUSY particles with 
HDECAY 5.10~\cite{Djouadi:1997yw} and SDECAY~\cite{Muhlleitner:2003vg}, respectively. Particularly relevant to this study are the 
calculations of the neutralino scattering cross sections and relic density, performed with micrOMEGAs~\cite{Belanger:2008sj} and 
SuperIso Relic v3.2~\cite{superiso,superiso_relic}, respectively. A total of $\sim 10^{7}$ valid pMSSM points have been simulated. Of these, 
575k are accepted after imposing the constraints from electro-weak, flavour, DM relic density and lower-energy data discussed in~\cite{Arbey:2012bp}. 
The relic DM density constraint is applied in the loose form by taking the upper limit from the PLANCK data~\cite{Ade:2013zuv}, but allowing 
other particles to contribute to the observed cosmic DM and/or modifications to the early universe properties, corresponding to 
$10^{-4} < \Omega_{\chi} h^2 < 0.163$. 
In order to test the compatibility of the accepted pMSSM points with the LHC searches, we simulate event samples and perform a parametric simulation 
for the event reconstruction. Events are generated with MadGraph 5~\cite{Alwall:2011uj} and Pythia 8.150~\cite{pythia8} with the CTEQ6L1 
parton distribution functions (PDFs)~\cite{Pumplin:2002vw} and the physics  observables are obtained from fast simulation performed using 
Delphes 3.0~\cite{Ovyn:2009tx}. Signal selection cuts are applied to simulated signal events, while the number of SM background events in the signal 
regions are from the experiment estimates. The 95\% confidence level (C.L.) exclusion of each SUSY point in presence of background 
only is determined using the CLs method~\cite{Read:2002hq}. 



The monojet searches are based on processes $pp \to \tilde{\chi} \tilde{\chi} + j$ which can be seen as a complementary process to the 
$\tilde{\chi} + p \to \tilde{\chi} + p$
scattering process of DM direct detection experiments, since they both probe the WIMP coupling with standard matter.
Monojets are a distinctive topology of events with a single, high $p_t$ hadronic jets, limited additional hadronic activity and large MET.  
Their relevance to the search of pair production of weakly-interacting, or non-interacting, particles was first exploited at the 
Tevatron~\cite{Chatrchyan:2012me} and they are now actively searched for at the LHC by ATLAS~\cite{ATLAS-CONF-2013-068} and CMS~\cite{CMS-EXO-12-048}, 
with the analysis of $\sim$20~fb$^{-1}$ of data at 8~TeV. 
The bounds from monojet searches can be interpreted as exclusion contours of the spin-independent WIMP scattering cross section, 
$\sigma_{\tilde{\chi} p}^{SI}$, as a function of the WIMP mass to compare with DM direct searches, relying on an effective approach 
with contact operators~\cite{Goodman:2010yf,Goodman:2010ku}. However, its validity is limited to the cases where a single heavy particle 
mediates the WIMP scattering with the nucleons. While this condition 
applies to simple DM models involving just one heavy mediator and a single new particle in the final states, it is generally not applicable 
to more complex models with several particles in the interaction, such as SUSY. 
In the specific case of the MSSM with neutralino LSP studied here, monojet events originate from the channel 
$pp \to \tilde{\chi}^0_1 \tilde{\chi}^0_1 + j$, mediated at tree level by $Z$ and Higgs bosons in s-channel and SUSY particles in 
t-channel, but also from processes such as $pp \to \tilde{q} \bar{\tilde{q}} + j$ or $pp \to \tilde{g} \tilde{g} + j$, in particular if 
SUSY particles in the final state decay to a $\tilde{\chi}$ and a soft quark or gluon. Such events are of importance when the splitting between 
the mass of the SUSY strongly-interacting particles with the LSP is small and make monojet searches especially valuable and complementary 
to the jets+MET channels in these scenarios. The impact of the monojet searches at Tevatron and the LHC in the MSSM has 
already been studied in simple, or simplified, models where only one scalar quark or gluino is close in mass to the lightest 
neutralino~\cite{Gunion:1999jr,Alwall:2008ve,Carena:2008mj,Alwall:2008va,Izaguirre:2010nj,LeCompte:2011cn,Ajaib:2011hs,LeCompte:2011fh,He:2011tp,Drees:2012dd,Dreiner:2012gx,Bhattacherjee:2012mz,Yu:2012kj,Dreiner:2012sh,Han:2013usa}.
In order to study the monojet searches in the more general cases offered by the pMSSM, where many SUSY strongly-interacting particles can have 
similar mass values, we simulate monojet events with MadGraph~5, generating the full $2\to3$ matrix elements for all the possible processes 
$pp \to \tilde{g}/\tilde{q}/\tilde{\chi}^0_1 + \tilde{g}/\tilde{q}/\tilde{\chi}^0_1 + j$, interfaced with the Pythia~6 parton 
shower~\cite{Sjostrand:2006za}. This method, specifically developed for our study, appears appropriate for the interpretation of monojet 
results in the context of general SUSY models. Because of computing time limitations, we generate the processes involving the LSP 
neutralino, the lightest of the SUSY strongly interacting particles (LSSP) and additional channels involving the other SUSY particles close 
in mass to the LSP or the LSSP. We verify that the inclusion of the heavier SUSY states does not significantly modify our results, on a selected 
set of pMSSM points, where all SUSY processes are generated.



MET searches at the LHC have considered a wide variety of topologies and final states. Here, we consider the preliminary results by the ATLAS 
Collaboration for the search of scalar quarks of the first two generations and gluinos in the jets+MET channel~\cite{ATLAS-CONF-2013-047}, 
of scalar top and bottom quarks with $b$-tagged jets and MET~\cite{ATLAS-CONF-2013-053} and chargino/neutralino associate production in the 
two-~\cite{ATLAS-CONF-2013-049} and three-leptons+MET channels~\cite{ATLAS-CONF-2013-035}.


Searches for WIMP neutralinos scattering in direct DM detection experiments probe the same physical vertex as some of the LHC searches. This 
makes the comparison of the reach and implications of these searches in the pMSSM particularly interesting. Several experiments have recently 
reported results corresponding to sensitivities relevant to the study of MSSM scenarios. The most recent result, from LUX, a dual-phase 
Xenon time projection chamber underground experiment~\cite{Akerib:2013tjd}, reported a minimum value of the 95\% C.L. upper limit of 
$\sigma_{\tilde{\chi} p}^{SI} = 10^{-9}$~pb for a WIMP mass of $\sim$40~GeV. 
The typical range for MSSM points is 10$^{-14} < \sigma_{\tilde{\chi} p}^{SI} <$ 10$^{-6}$~pb for 
100$< M_{\tilde{\chi}} <$1000~GeV~\cite{Arbey:2012bp}. 
The Xenon-100~\cite{Aprile:2012nq} and LUX results challenge the excesses of events reported by several other 
experiments~\cite{Bernabei:2010mq,Aalseth:2011wp,Angloher:2011uu,Ahmed:2010wy}, which may be interpreted as due to a low-mass WIMP with 
large scattering cross section. We consider the effect of these bounds on the pMSSM parameters by imposing the 95\% C.L.\ upper limit on 
$\sigma_{\tilde{\chi} p}^{SI}$ obtained by LUX on our scan points. Similar results are obtained using the Xenon-100 limit.

\begin{figure}[h!]
\vspace*{-0.425cm}
\begin{tabular}{cc}
\includegraphics[width=0.525\columnwidth]{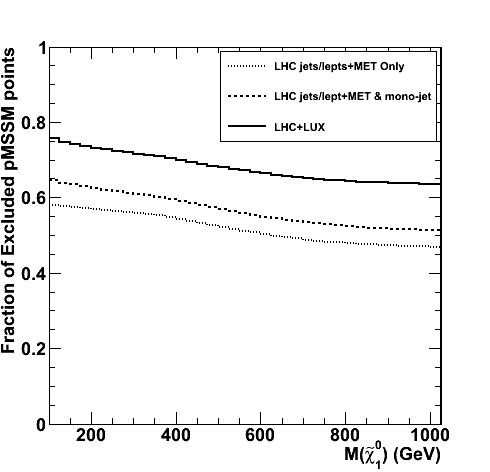} &
\hspace*{-0.25cm}\includegraphics[width=0.525\columnwidth]{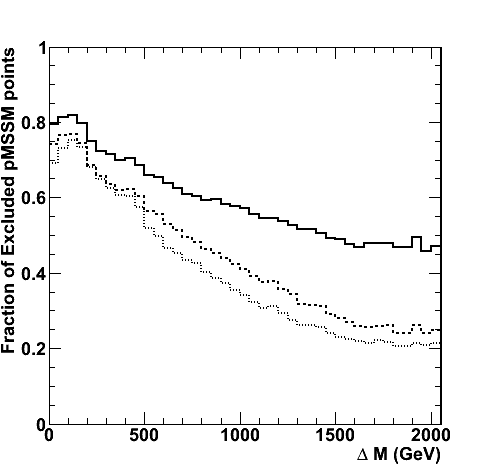} 
\end{tabular}
\vspace*{-0.425cm}
\caption{Fraction of pMSSM points excluded by the jets/leptons+MET (dotted line), their combination with the monojet analyses (dashed line) and 
that with the LUX direct DM search (continuous line) as a function of $M_{\tilde{\chi}}$ (left panel) and $\Delta M$ (right panel).} 
\label{fig:1D}
\end{figure}

The sensitivity of the searches discussed above can be quantified first by studying the fraction of accepted pMSSM points incompatible with a 
given channel or combination of channels at 95\% C.L. Results are summarised in Table~I.
\begin{table}[h!]
\caption{Fraction of pMSSM points rejected by the various LHC channels and the direct DM searches.}
\begin{tabular}{|l|c|c|}
\hline
Search & \multicolumn{2}{|c|}{Fraction of pMSSM Points} \\
\hline
       & Excluded                 & Excluded Uniquely  \\
\hline
Jets+MET  & 0.455 & \\
$b$-Jets+MET  & 0.178 & \\
$\ell$s+MET & 0.017 & \\ \hline
All Jets/$\ell$s+MET  & 0.470 & \\ \hline
monojet+MET & 0.262 & 0.045 \\
All LHC MET Searches & 0.515 & 0.300 \\ \hline
DM Searches & 0.335 & 0.120 \\ \hline
All Searches & 0.635 & - \\
\hline
\end{tabular}
\label{tab:stat}
\end{table}
About two-third of the accepted pMSSM points, compatible with the loose $\Omega h^2$ relic density constraint, are rejected by the combination 
of the LHC and LUX bounds considered here. As anticipated, the inclusion of the monojet and DM direct search ensures the exclusion of a 
significant fraction of points, not excluded by the jets/leptons+MET LHC searches, corresponding to 17\% of our accepted pMSSM points. 
Contrary to the effective operator approach that would predict only of order $2 \times 10^{-4}$ of the points to be excluded by the monojet 
bounds, the analysis of the $2\to3$ SUSY processes for our pMSSM points gives that 26\% of these are rejected by the LHC monojet analyses.
\begin{figure}[ht!]
\vspace*{-0.425cm}
\begin{tabular}{cc}
\includegraphics[width=0.475\columnwidth]{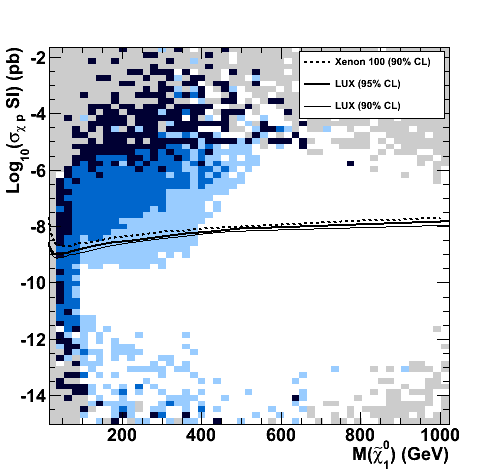} &
\includegraphics[width=0.475\columnwidth]{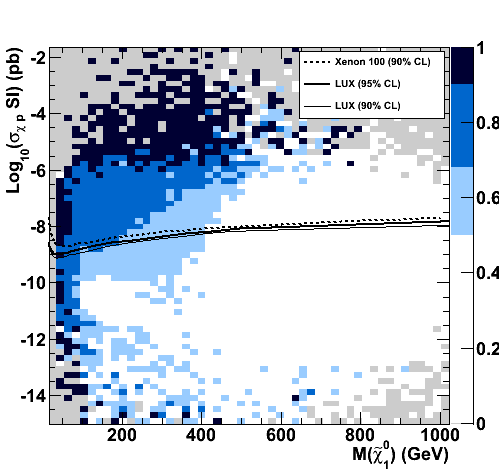} \\
\end{tabular}
\vspace*{-0.45cm}
\caption{Fraction of pMSSM points excluded by jets/leptons+MET (left)
and their combination with monojet searches (right) 
in the $\sigma_{\tilde{\chi} p}^{SI}$ - $M_{\tilde{\chi}}$ parameter plane. 
The upper limits from direct detection experiments are also indicated.}
\label{fig:sigmaMN1}
\end{figure}
Then, we study our results in terms of the fraction of accepted pMSSM points in our scans, excluded 
by the jets/leptons/monojet+MET LHC and the direct DM searches discussed above, as a function of a single, or a pair of, pMSSM parameter to study the 
parameter regions which benefit from the inclusion of the monojet and direct DM searches. In the context of this study the parameters of interest are the 
mass of the WIMP candidate, the neutralino LSP, $M_{\tilde{\chi}^0}$, the LSSP mass, $M_{\tilde{q},\tilde{g}}$ 
and the mass splitting,  $\Delta M = M_{\tilde{q},\tilde{g}} - M_{\tilde{\chi}^0}$. In general, the masses determine the production cross section at the LHC, 
while $\Delta M$ controls the signal event kinematics and thus the selection efficiency. 
In particular, we are interested in assessing the exclusion reach of the LHC and direct DM searches in the $\sigma_{\tilde{\chi} p}^{SI}$ - $M_{\tilde{\chi}}$ 
and $M_{\tilde{q},\tilde{g}}$ - $M_{\tilde{\chi}}$ planes and contrast the sensitivity of the various channels, identifying specific regions of the parameter 
space which are unique to the monojet analyses and the direct DM searches. First we consider the parameter plane  
$\sigma_{\tilde{\chi} p}^{SI}$ - $M_{\tilde{\chi}}$ where the comparison to the bounds from direct detection experiments is straightforward 
(see Figure~\ref{fig:sigmaMN1}). The sensitivity of the LHC monojet searches provides a bound which extend to larger WIMP-$\tilde{\chi}$ masses along a line 
at almost constant scattering cross section compared to the jets+MET analyses. The analysis of the $M_{\tilde{q},\tilde{g}}$ - $M_{\tilde{\chi}}$ plane provides 
us with a good illustration of the complementarity of the jets/lepton+MET, monojet and direct DM searches. 
\begin{figure}[h!]
\vspace*{-0.425cm}
\includegraphics[width=0.75\columnwidth]{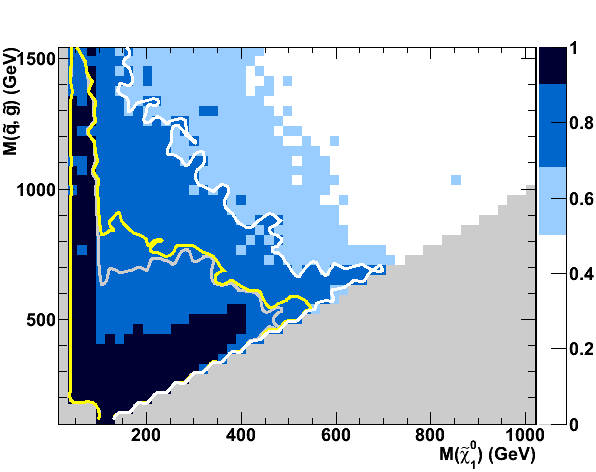} 
\vspace*{-0.425cm}
\caption{Fraction of pMSSM points excluded by the combination of the LHC jets/leptons+MET, monojet analyses and direct DM 
searches in the $M_{\tilde{q},\tilde{g}}$ - $M_{\tilde{\chi}}$ parameter plane. The lines give the parameter region where 
68\% of the pMSSM points are excluded by the jets/leptons+MET searches alone (grey line), the combination with monojet searches 
(yellow line) and also with the direct DM LUX experiment (white line).}
\label{fig:MSqMN1}
\end{figure}
Figure~\ref{fig:MSqMN1} shows the fraction of points excluded at 95\% C.L. by the combination of all these searches and also the contours 
enclosing the region where 68\% of the pMSSM points are rejected by jets/lepton+MET channels and their combination with monojets. The latter 
channel, with its specific sensitivity to low-mass SUSY particles, in particular in degenerate scenarios due to the build-up of the monojet  
cross section, significantly improves the rejection of points along the $M_{\tilde{q},\tilde{g}} \simeq M_{\tilde{\chi}}$ line, where the jets+MET 
searches are weaker, due to the reduced transverse momentum of the hadronic jets. This effect shows in a gain of $\sim$80~GeV in sensitivity to 
the neutralino LSP mass close to the limit $M_{\tilde{q},\tilde{g}} \simeq M_{\tilde{\chi}}$ and a 15\% wider surface covered by the contour. 
The gain is further enhanced with the inclusion of the LUX bound by $\sim$120~GeV in sensitivity to the WIMP $\tilde{\chi}$ mass and 62\% wider surface. 
The extent of these gains reduces when requiring larger fractions of pMSSM points to be rejected. The surface covered by the contour enclosing 
the region with 90\% of the pMSSM points rejected increases by only 8\% with the addition of the monojets and by 30\% also with the LUX bounds.
\begin{figure}[h!]
\vspace*{-0.425cm}
\includegraphics[width=0.75\columnwidth]{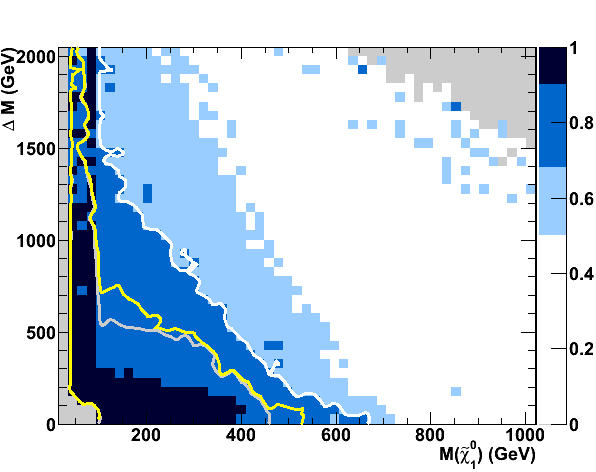} 
\vspace*{-0.45cm}
\caption{Fraction of pMSSM points excluded by the combination of the LHC jets/leptons+MET, monojet analyses and direct DM 
searches in the  $\Delta M$ - $M_{\tilde{\chi}}$ parameter plane. The conventions for the lines are as in Figure~\ref{fig:MSqMN1}.}
\label{fig:DMMN1}
\end{figure}
The fraction of pMSSM points excluded at 95\% C.L. by the LHC and DM searches is shown as a function of the $\tilde{\chi}$ mass and $\Delta M$, when 
integrating over all other pMSSM parameters in Figure~\ref{fig:1D}. The distribution is rather smooth in the $\tilde{\chi}$ mass value but the 
increase in sensitivity at very small and intermediate- to large-$\Delta M$ values afforded by the inclusion of the monojet and DM search limits is 
evident. 
Finally, we estimate the effect of systematic uncertainties by varying the production cross sections of $pp \rightarrow \tilde{q} \tilde{q}$, 
$\tilde{g} \tilde{g}$ by $\pm$20\% to account for PDF uncertainties~\cite{Nadolsky:2008zw}, those of 
$pp \to \tilde{\chi}^0_1 \tilde{\chi}^0_1 + j$, $\tilde{q} \tilde{q} + j$ by $\pm$30\% to include also uncertainties from the 
generation~\cite{Dreiner:2012sh}. However, their effects are small. The $\sigma_{\tilde{\chi} p}^{SI}$ bounds from direct detection 
DM experiments depends on assumptions on the local DM density in the Galaxy. We propagate the uncertainty on the local DM density 
$\rho = (0.3 \pm 0.1)$~GeV/cm$^3$~\cite{Bovy:2012tw} to the $\sigma_{\tilde{\chi} p}^{SI}$ 95\% C.L. bound.
Their effect in the $M_{\tilde{q},\tilde{g}}$ - $M_{\tilde{\chi}}$ parameter plane are summarised in Figure~\ref{fig:syst}, 
which shows that the qualitative results of our study remain valid after accounting for these theoretical uncertainties.
\begin{figure}[h!]
\vspace*{-0.425cm}
\includegraphics[width=0.75\columnwidth]{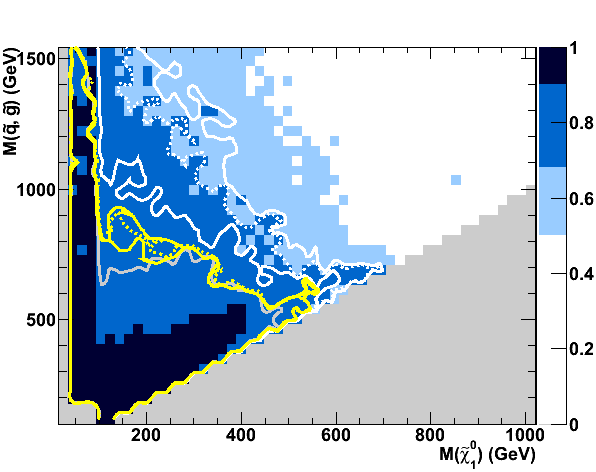} 
\vspace*{-0.45cm}
\caption{Ranges of the contours as given in Figure~\ref{fig:MSqMN1} obtained by modifying the production and the scattering cross sections to 
account for systematic uncertainties, as discussed in the text. The contours corresponding to the values used for Figure~\ref{fig:MSqMN1} 
are given by the dotted lines.}
\label{fig:syst}
\end{figure}


{\it The authors are grateful to several colleagues for discussion on the topics presented in this letter. 
In particular, R.~Gaitskell and J.~Chapman provided the LUX results in numerical form, B.~Fuks advised 
on the monojet cross section calculations and C.~Wagner engaged in useful discussion. We also acknowledge 
the LPCC and the CERN PH-TH unit for computing support.
The work of A.A.\ was supported by the F\'ed\'eration de Recherche
A.-M. Amp\`ere de Lyon; that of F.M., in part, by the National Science Foundation under Grant No. PHYS-1066293 
and profited of the hospitality of the Aspen Center for Physics.}



\end{document}